\documentclass[12pt]{article}
\usepackage{setspace}
\usepackage{endnotes}
\let\footnote=\endnote
\usepackage{sectsty}
\subsectionfont{\normalfont\itshape}

\usepackage{amsmath}
\usepackage{amssymb}
\usepackage{amsthm}
\usepackage{mathrsfs}
\usepackage{authblk}
\usepackage[title]{appendix}
\usepackage{tikz}
\usepackage{multirow}
\usepackage{multicol}
\usepackage{mathrsfs}
\usepackage{cite}
\usepackage{hyperref}
\usepackage[round]{natbib}
\usepackage{multibib}
\newcites{sec}{References for the Appendices}
\usepackage{fullpage}
\usepackage[export]{adjustbox}
\usepackage{subcaption}
\newcommand{\specialcell}[2][c]{%
  \begin{tabular}[#1]{@{}c@{}}#2\end{tabular}}
\title{Identifying Films with Noir Characteristics Using Audience's Tags on MovieLens}
\author[ \hspace{-1ex}]{Ziyue Zhu}
\affil[ \hspace{-1ex}]{School of Statistics, University of Minnesota}
\affil[ \hspace{-1ex}]{Correspondence: 224 Church Street SE, Minneapolis, MN 55455, USA Email: zhux0502@umn.edu}
\date{}
\begin{document}
\maketitle
\begin{abstract}
\noindent 
We consider the noir classification problem by exploring noir attributes and what films are likely to be regarded as noirish from the perspective of a wide Internet audience. We use a dataset consisting of more than 30,000 films with relevant tags added by users of MovieLens, a web-based recommendation system. Based on this data, we develop a statistical model to identify films with noir characteristics using these free-form tags. After retrieving information for describing films from tags, we implement a one-class nearest neighbors algorithm to recognize noirish films by learning from IMDb-labeled noirs. Our analysis evidences film noirs' close relationship with German Expressionism, French Poetic Realism, British thrillers, and American pre-code crime pictures, revealing the similarities and differences between neo noirs after 1960 and noirs in the classic period.
\end{abstract}
\section{Introduction} \label{intro} 

We use audience-provided tags from MovieLens, a web-based recommendation system, to develop a machine-learning algorithm to identify films that are noirish and hence provide insight into what potential noir qualities such films may bear. While machine-learning algorithms have been applied to film classifications and genre studies, little has been done on film noir. This may be partly due to the debate on what noir is (\citealp{durgnat1970family}; \citealp{schrader1972notes}; \citealp{naremore2008more}; \citealp{spicer2013companion}). Instead of attempting to define film noir precisely, we will focus on the boundary of the noir category in the same spirit as the more traditional approach of  \citet{naremore2008more}. We concentrate on a broader audience's characterization of films and use it to recognize film noirs and their cousins. 

The word, `noir,' was applied by the French critics in the 1940s to describe a number of American criminal films often adapted from hard-boiled novels \citep{chartier1946americains, frank1946nouveau}. These films include John Huston's \textit{The Maltese Falcon} and Billy Wilder's \textit{Double Indemnity}, which are frequently cited as noir masterpieces. Although the specific term, `film noir,' went unnoticed in the US at that time, interest in it grew eventually thanks to the film critics and enthusiasts \citep{naremore2008more}.

\citet{borde2002panorama} point out that `the presence of crime' is noirs' `most distinctive stamp,' and that film noir `is set in the criminal milieu itself and describes the latter,' in many cases centered on the ambiguous protagonist typified by Humphrey Bogart and the ambiguous femme fatale `who is fatal unto herself.' 
\citet{place1974some} focus on the visual motifs of noir, talking about the antitraditional photography and mise-en-scène, including low-key lighting, greater depth of field and imbalanced composition. \citet{porfirio1976no} indicates, however, that neither the presence of crime nor the visual style captures all the noir traits. He argues that it is `an existential attitude towards life' that unifies diverse film noirs.

Is film noir a genre at all? \citet{durgnat1970family} believes it is and divides it into eleven groups.
\citet{schrader1972notes}, on the other hand, considers noir as `a specific period of film history' which started from \textit{The Maltese Falcon} and ended with Orson Welles' \textit{Touch of Evil}, saying noir is not a genre as it is not defined `by conventions of setting and conflict.' He examines the emotions and emphasizes the tone and mood of noirs, taking into account the cultural background, such as post-war realism. However, `it has always been easier to recognize a film noir than to define the term' \citep{naremore2008more}.

Scholars and critics agree that a couple of films are noir.  \citet{maltby1984film}, for instance, conceives Jacques Tourneur's classic, \textit{Out of the Past} as one of them: `Whatever film noir is, \textit{Out of the Past} is undoubtedly film noir.' Yet seeking an exhaustive list of noirs is challenging, as one would need to examine every film in detail. Thus, most analyses of noir are based on small sample sizes. This issue is not confined to the essay by \citet{place1974some} as \citet{naremore2008more} observes, but is commonly the case when film scholars discuss the noir category. An essential noir like \textit{Double Indemnity} has been studied by numerous scholars, but the film industry is replete with many more productions that have gained much less attention.

Online tagging systems, which have flourished since 2004 \citep{hammond2005social,sen2006tagging}, allow the audience to add tags to films. Studying a large amount of films is thus made possible by investigating tags applied by a wide audience via machine learning. We analyze tags for describing films added by users of MovieLens, a web-based recommendation system, to recognize films with noir characteristics. 

Although tags do not directly extract information from films in the same way visual features do, they are less expensive and less time-consuming to obtain. Computable video features, such as average shot length, color variance, visual disturbance, among others \citep{fischer1995automatic, rasheed2002movie, rasheed2005use}, as well as some text information, such as closed captions \citep{brezeale2006using}, have been used in many of the machine learning approaches to classifying films. While these features are useful for content analysis, acquiring them can be difficult and expensive, for one has to encode all films analyzed or at least their previews.

\begin{figure}
     \centering
     \begin{subfigure}[b]{0.9\textwidth}
         \centering
         \includegraphics[width=\textwidth]{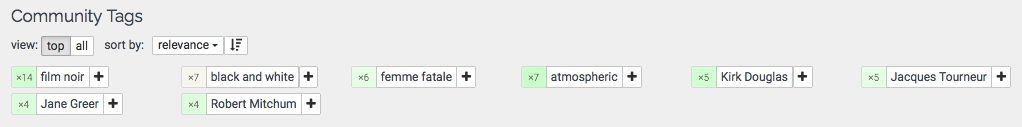}
         \caption{Tags of \textit{Out of the Past}}
     \end{subfigure}
     \hfill
     \begin{subfigure}[b]{0.9\textwidth}
         \centering
         \includegraphics[width=\textwidth]{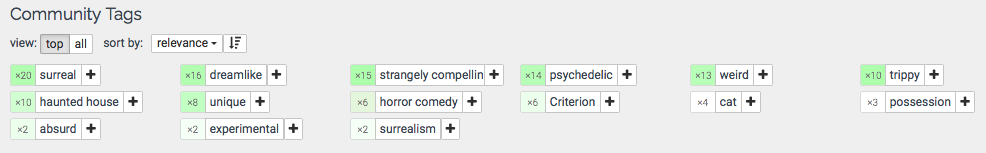}
         \caption{Tags of \textit{House}}
     \end{subfigure}
        \caption{Community tags of Jacques Tourneur's classic noir \textit{Out of the Past}, and Nobuhiko Ôbayashi's cult horror \textit{House} on MovieLens} 
        \label{tagseg}
\end{figure}

Compared with other less expensive datasets such as film plots, tags cover more kinds of information. Plots have been used to classify personas of film characters \citep{bamman2013learning}. But for the classification of films themselves, the lack of the description of a film's style  in its plot could easily lead to incorrect results. Tags, on the other hand, range from names of directors and actors, and personas of characters, to the audience's impression of the whole film. Figure \ref{tagseg} gives screenshots showing frequent community tags of the classic noir \textit{Out of the Past}\footnote{https://movielens.org/movies/2066} and Nobuhiko Ôbayashi's cult horror \textit{House}.\footnote{https://movielens.org/movies/50641} Not surprisingly, the audience says the former is `atmospheric' and the latter is `strangely compelling' and `surreal,' which are characteristics that might not be reflected in plot summaries. 

Compared to written reviews that allow free-form descriptions, tagging systems let users add their own and see others' keywords directly. It is less time-consuming for users, and as more and more of them add their own tags, a dataset that is larger and more interpretable is naturally produced.

As described in Section \ref{model}, we apply a one-class nearest neighbors classification algorithm to approach this problem. We identify films with noir traits according to audience's tags via calculating their distances to those classified as noir by IMDb, which are assumed to be the only known noirish films. The results in Section \ref{discussion} reveal the nexus between noirs and German Expressionism, French Poetic Realism, British thrillers as well as the American pre-code crime pictures, and implies how neo noirs after 1960 are similar to and at the same time diverge from noirs in the classic period.

\section{Dataset Description} \label{db} 
\subsection{Narrative films for analysis}
We  begin by introducing the datasets and the procedure for pre-processing the free-form tags. We mainly explored the tagging data in MovieLens\footnote{We use the 25M dataset on https://grouplens.org/datasets/movielens/} and also use films' factual information from the IMDb dataset.\footnote{https://www.imdb.com/interfaces/}  

\begin{figure}
     \centering
         \includegraphics[width=0.75 \textwidth]{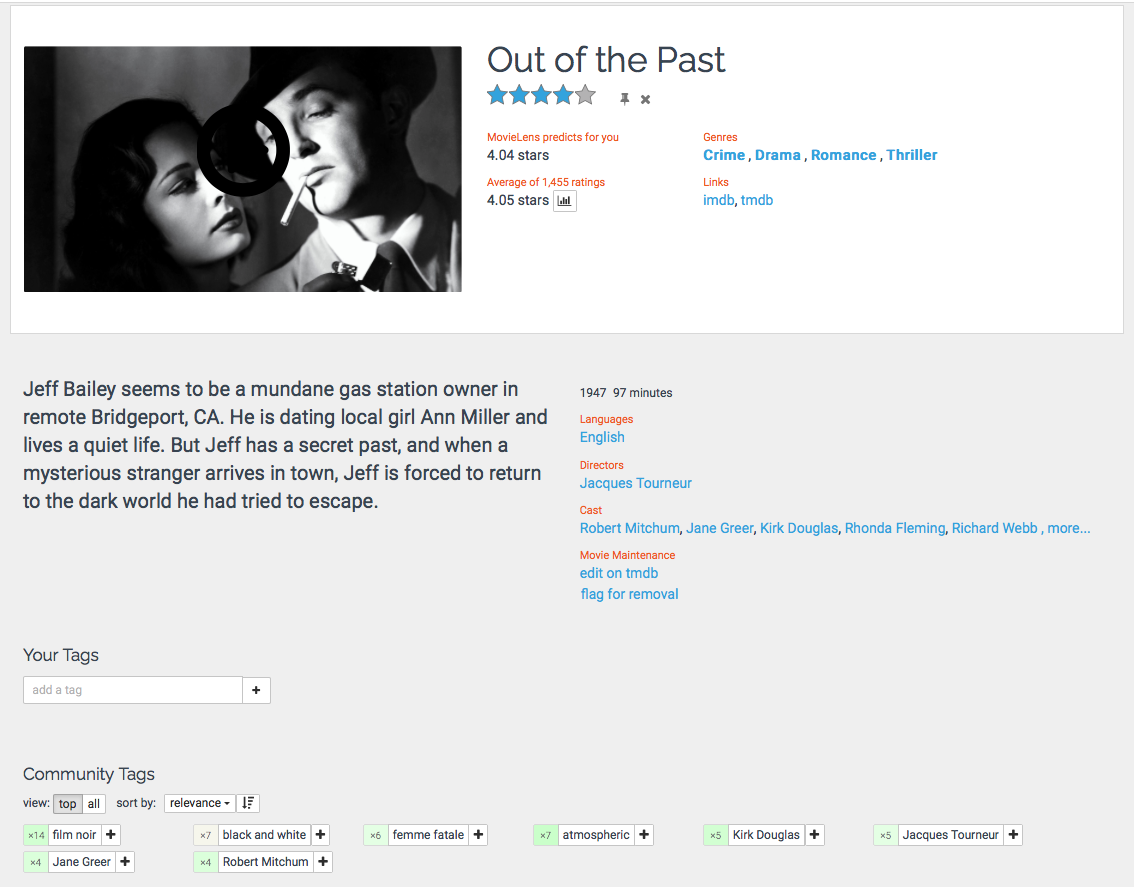}
         \caption{The MovieLens webpage of \textit{Out of the Past}}
        \label{oop}
\end{figure}

While users can rate and review films on influential online platforms such as IMDb and Amazon Prime Video, they are not allowed to add tags. MovieLens, which was released in 1998, on the other hand, introduced a tagging feature in late 2005 and forms a database of ratings and tags \citep{harper2015movielens}. Figure \ref{oop}, the screenshot of \textit{Out of the Past}'s MovieLens webpage, shows that the users can either write their own tags or choose the commonly-used community tags to describe a film. There are 62,423 film titles in the MovieLens dataset that was last updated in November 2019. The file links.csv provides the unique MovieLens ID for each title as well as its IMDb IDs, and the file tags.csv gives 1,093,360 tags (73,051 of them are different) added by 14,592 users for the films in the database. The users' demographic data are not provided. The IMDb dataset downloaded on Jan 21, 2020 has the file title.basics containing information of release year, runtime and genre for 6,484,013 films. Since we focus on the audience's tags, we will only consider the films existing in the MovieLens dataset, which tend to be the more popular films among the audience. 

In the links.csv file, 74 IMDb IDs are outdated and thus could not be used to find their IMDb information. We update these 74 IDs manually, exluding one film that is not currently in the IMDb database. Furthermore, of the 62,422 film titles that exist in both the MovieLens and IMDb datasets, there are 25 pairs of duplicated IMDb IDs, meaning that 25 films have two distinct IDs each on MovieLens. Therefore, we will identify films according to their IMDb IDs instead of their MovieLens IDs.

The IMDb dataset classifies film titles into 10 types: short, movie, tvMovie, tvSeries, tvEpisode, tvShort, tvMiniSeries, tvSpecial, video, and videoGame. We are only interested in the types movie and tvMovie, making a total of 56,630 titles. Furthermore, we only consider narrative films, so we take away films classified as documentary by IMDb, leaving 51,402 films.  
\subsection{Pre-processing the tagging data}
Not all films have tags. Of the 51,402 narrative films of IMDb title types movie and tvMovie in our dataset, 38,366 of them are tagged in MovieLens. Quantifying the relationship between tags and films has been explored by \citet{vig2012tag}, who computed $1,128$ common tags
relevance scores for films. However, they failed to consider the similarity between the tags themselves. For example, it doesn't make sense for `zombie' and `zombies' to have different relevance scores for the same film. Additionally, when they select the tags based on the number of users applying them at the beginning, 
some useful tags are deleted only due to different spellings. For instance, a minority of users use `noire' instead of `noir,' although they are referring to the same thing; some type `funny!' instead of `funny.'

To solve such problems, we first pre-process the tags. If two tags differ from each other only by punctuations, extra white spaces, or capitalization, then they are considered to be the same tag. We then get the stems of the tags by removing the inflected parts of every word in every tag. We have $57,168$ different stems in total. For simplicity, in the following sections, we call these stems `tags,' and a stem is referred to by the tag it is from. If there are multiple tags sharing the same stem, then one of them is used for representing it. For example, the stem `zombi' comes from `zombie' or `zombies,' and when talking about this stem, we simply write `zombie.' One of the stems, `heroin' may denote two words of totally different meanings: heroine and heroin. We define our own stem `heroine' and manually assign the tag `heroine' to it and assign only the tag `heroin' to the stem `heroin.' 

Similar to \citet{vig2012tag}, we first remove tags that have been applied by 10 or fewer users as these tags are usually too personal to characterize the films. We also remove tags that have been applied to no more than 10 films, as they only represent a minority of films. Next, tags related to people's names, including directors', screenwriters', actors' and actresses', and any other real names are removed. These do not cover the names of fictional characters, such as `James Bond' and `Sherlock Holmes.' We check tags varying only in the usage of white spaces, and there are 7 pairs of them, such as `anti hero' and `antihero,' `art house' and `arthouse.' Each pair is regarded as a single tag. 

After this pre-processing, we have $2,862$ tags and $35,049$ films with at least one of them. Among the tags, 74 of them carry no useful information for film classification, so they are removed. Some of these are personal labels such as `in Netflix queue' and `to watch again.' Others are factual, including the films' production company information such as `Pixel,' DVD distributor such as `Criterion' (short for Criterion Collection), the lists films enter such as `AFI 100' (short for American Film Institute's `100 Years... 100 Films'), and the awards and nominations gained such as `Oscar' and `Palme d'Or.' Not all of the factual tags, however, are removed. For example, `AFI 100 laugh,' namely, being on the AFI's `100 Years... 100 Laughs' list, at least indicates the film is funny according to AFI's criteria.\footnote{https://www.afi.com/afis-100-years-100-laughs/} So we keep such tags. After these tags are removed, some films do not have tags anymore; there are now $N=$34,412 films, and $L$=2,788 tags left for analysis. 
\section{One-class nearest neighbors algorithm for film noirs}\label{model}
\subsection{Methodology overview}
A one-class classification algorithm to predict if a film has noir characteristics using MovieLens tags is applied. Given examples of film noirs, we are to determine whether a film with tags on MovieLens has noir characteristics from the audience's viewpoint.

More specifically, films are assumed to belong to either the positive class $\mathcal{P}$, denoting those with noir characteristics, and the negative class, denoting those without. The training set $\mathcal{T}$ used to build the model contains only objects in $\mathcal{P}$, which are the films classified as noirs by IMDb. All the other films are in the unlabeled set $\mathcal{U}$, meaning to which class they belong is yet to be known. Our task is to use MovieLens tags for films applied by the website's users to predict which objects in $\mathcal{U}$ are in $\mathcal{P}$. There are $|\mathcal{T}|=453$ IMDb-labeled noirs and $|\mathcal{U}|=33,959$ unlabeled films in our dataset.

We perform the one-class classification because there are only positive objects in our training set. Negative training objects cannot be included for two reasons. First, obtaining non-noir films in the IMDb dataset is not feasible, for it only tells which films are noirs, but not which are not noirs. IMDb claims that it takes the view that film noir `began with \textit{Underworld} (1927) and ended with \textit{Touch of Evil} (1958),'\footnote{See footnote 4.} indicating that it does not identify any noirs before 1927 or after 1958. Whether it recognizes all noirs between 1927 and 1958 is also doubtful: for example, John M. Stahl's work \textit{Leave Her To Heaven} is not labeled as noir by IMDb, but is often considered as noir among film scholars \citep{borde2002panorama, naremore2008more, selby1984dark}. Additionally, IMDb lists up to three genres for each film.\footnote{See footnote 4.} So if a film does not have an IMDb noir label, it does not mean it is in the negative class. Second, even if we utilize other film datasets or study various academic essays on noir, collecting a sample of convincing and representative non-noirs is impractical. While noir filmographies can be found in a number of publications such as \citet{borde2002panorama}, \citet{ward1979film} and \citet{selby1984dark}, a list of non-noirs is much less common. Given the variety of non-noirs, making a list typical of them will be arduous to say the least, if not impossible. Consequently, there are no negative objects, namely, non-noirs, that can be put in the training set. In this setting, the labeled training set, $\mathcal{T}$, IMDb noirs, are only from the positive class, $\mathcal{P}$, and all other films are considered as unlabeled data, $\mathcal{U}$.

We further illustrate why the film category information from IMDb is used to form our training set rather than other sources. An influential online film dataset is preferred, for we focus on the behavior of the internet audience and want to train our model based upon the data they often search for film information from and tend to trust. IMDb becomes a reasonable option, as it is a large and popular database placing films into 28 `genres' including film noir.\footnote{https://help.imdb.com/article/contribution/titles/genres/GZDRMS6R742JRGAG} Here we are not concluding noir should be a genre instead of, say, a style, but simply treating the so-called IMDb `genre' as a category of films. Since MovieLens itself also lists films' genres provided by its web developers, one might think it would be more natural to use noirs there as our training objects. However, the classification of noirs by IMDb is more consistent than MovieLens as it clearly states what view it takes. As complicated as the `noir or not' question is, MovieLens, on the other hand, fails to give an explanation of how it decides whether a film is a noir.  \\
For all the films in $\mathcal{U}$, we want to quantify their `distance' to $\mathcal{T}$ and decide whether they should belong to $\mathcal{P}$, or namely, whether they have noir characteristics, via the nearest neighbors method. Given a film $z$, its nearest neighbors in $\mathcal{T}$ refer to the IMDb noirs that are most similar to it. We consider $z$'s multiple nearest neighbors, $NN_1(z),NN_2(z),...,NN_J(z)$ and then the nearest neighbor of these $NN_j(z)$'s, $NN(NN_j(z))$. The distances between $z$ and $NN_j(z)$ as well as the ratio of the mean of distances between $z$ and $NN_j(z)$ to
the mean of distances between $NN_j(z)$ and $NN(NN_j(z))$ are examined to give the result.

To calculate the distances between films, we use vectors to represent films. Each film can be assigned a binary vector, each coordinate of which corresponds to a tag, and is equal to $1$ if at least one user applies this tag to the film, and $0$ otherwise. However, tags might be related to each other. If we directly use the binary vectors regarding tag use for films for the one-class classification, we will be ignoring the similarity between films when their tags are different but of close meanings. Therefore, we cluster these tags first and then assign a feature-weighted vector to each film quantifying its relevance to the tag groups. Hence, our one-class classification includes two steps:\\
1) We assign a feature-weighted vector to each film based on its tags.\\
2) We perform one-class classification using the vectors obtained from the first step.\\
\subsection{Assigning feature-weighted vectors to films}
\label{ff}
After pre-processing the tags, we have partially dealt with synonymy and polysemy, the two issues often arising in semantic analysis \citep{deerwester1990indexing}. However, there still exist related tags. `Zombie' and `zombies' are now combined as one tag since they have the same stem, but tags like `France' and `French' are not. If classification is performed using the film-tag binary matrix directly, we will have correlated predictors. Suppose film A only has one tag `France,' film B only has `Paris,' and film C only has `thriller,' then any two of them do not share common tags. But actually A and B may be a more similar pair since Paris is capital of France. One remedy for such issue is to model the relationship between tags. But in this high-dimensional scenario, it would be computationally expensive to study the relationships between $L=2,788$ categorical variables.

To overcome this difficulty, we assume that there is a latent variable, tag group, to which the film and the tag are associated. A tag group consists of tags that are closely related to each other, and each film is then represented by a feature-weighted vector reflecting its connections with all these groups.

To figure out which tags are related, we observe the co-occurrences of tags by calculating the cosine similarity between them. A given tag is said to be strongly related to those tags with the highest cosine similarity to it. If tag $t_{l_1}$ is strongly related to tag $t_{l_2}$, we draw an edge from $t_{l_1}$ to $t_{l_2}$ with weight equal to the cosine similarity between these tags. All the tags and their edges thus form a directed graph, and a connected subgraph is a partition of it in which one can get from every tag to every other one through edges. In this way $478$ connected subgraphs are built, and for each of them, using the cluster\_optimal function in the igraph package in R \citep{csardi2006}, we divide them into smaller partitions, or namely, clusters. The technical details of the clustering analysis can be found in Appendix \ref{clustering}.

By choosing the cosine similarity, we avoid disadvantages of some other similarity measures. \citet{levenshtein1966binary} measures the tag similarity based on their spelling. This might work for tag pairs such as `wolf' and `wolves' (and note that for the word `wolf,' the singular and the plural form do not share the same stem), but not for `France' and `Paris.' \citet{begelman2006automated} examine the frequency counts of all the co-tag pairs and then find the `strongly related' tags for every tag using some cutoff point. However, this will lead to a bias towards the more frequently appearing tags. 

We cluster the tags using this machine learning algorithm instead of manually looking for groups of synonyms because the latter would be not only time-consuming but also subjective. For example, how do we know if `France' is more related to `Paris' or `French new wave'? How do we make sure that we can find all the tags related to `France'? 

\begin{figure}
     \centering
     \begin{subfigure}[b]{0.3\textwidth}
         \centering
         \includegraphics[width=\textwidth]{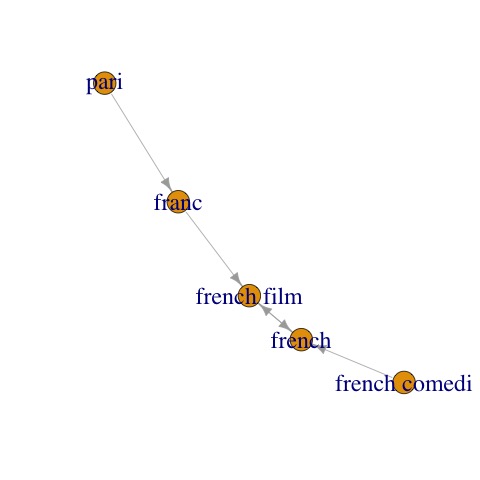}
         \caption{}
     \end{subfigure}
     \hfill
     \begin{subfigure}[b]{0.3\textwidth}
         \centering
         \includegraphics[width=\textwidth]{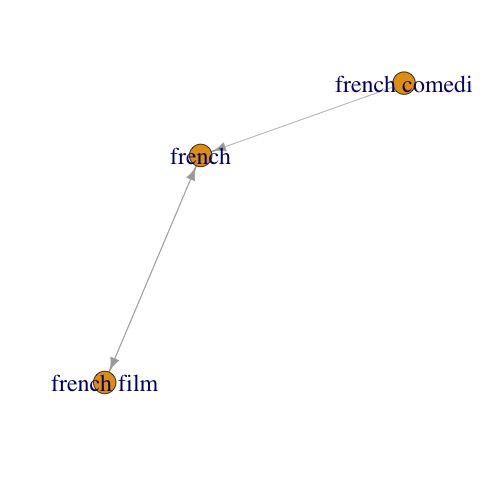}
         \caption{}
     \end{subfigure}
     \hfill
     \begin{subfigure}[b]{0.3\textwidth}
         \centering
         \includegraphics[width=\textwidth]{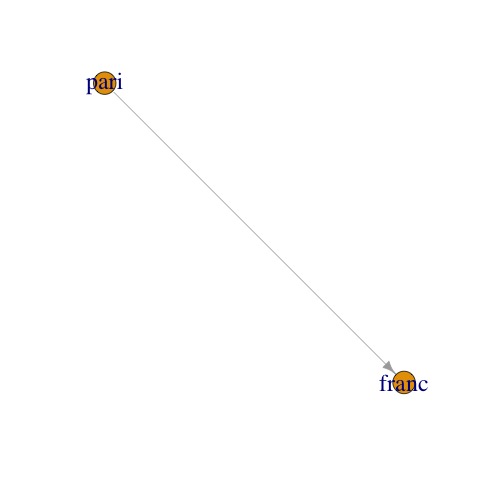}
         \caption{}
     \end{subfigure}
        \caption{The connected subgraph including the vertex `France' is displayed in (a). Note that `French New Wave' does not even appear here. The two partitions of the subgraph in (a) are displayed in (b) and (c). Each of these partitions forms a group of tag.}
        \label{paris}
\end{figure}

\begin{figure}
     \centering
     \begin{subfigure}[b]{0.3\textwidth}
         \centering
         \includegraphics[width=\textwidth]{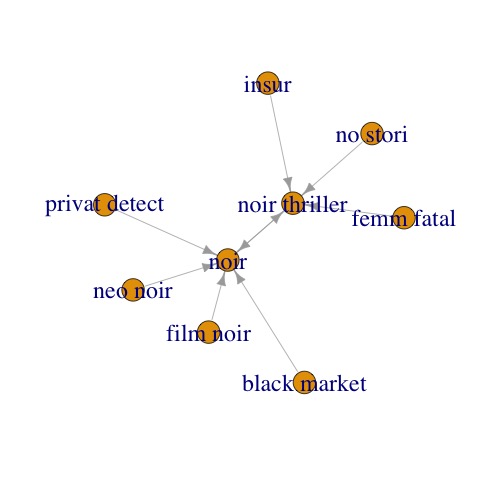}
         \caption{}
     \end{subfigure}
     \hfill
     \begin{subfigure}[b]{0.3\textwidth}
         \centering
         \includegraphics[width=\textwidth]{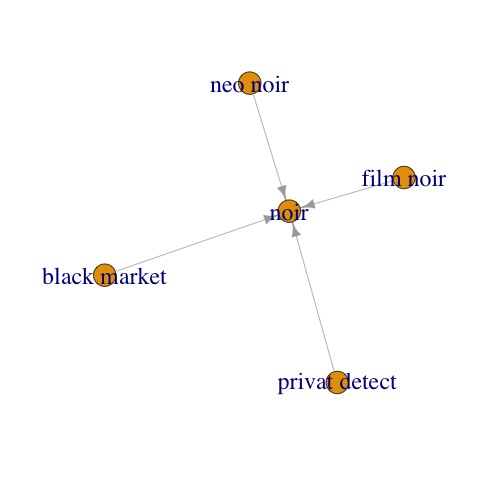}
         \caption{}
     \end{subfigure}
     \hfill
     \begin{subfigure}[b]{0.3\textwidth}
         \centering
         \includegraphics[width=\textwidth]{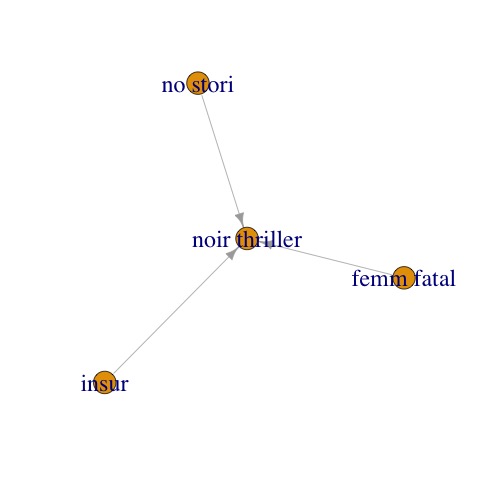}
         \caption{}
     \end{subfigure}
        \caption{The connected subgraph including the vertex `noir' is displayed in (a). The two partitions of the subgraph in (a) are displayed in (b) and (c). Each of these partitions forms a group of tags.}
        \label{noir}
\end{figure}

There are $M=$1,043 clusters of tags in total, with the smallest containing only one tag and the largest containing $19$ tags. Figure \ref{paris} shows the partitions of the connected subgraph \{Paris, France, French, French comedy, French film\}. `Paris' and `France' form one tag group while `French,' `French comedy' and `French film' form another. It may be because `Paris' and `France' are more likely to refer to the locations where the story takes place, and `French,' `French comedy' and `French film' are more related to the French cinema. Figure \ref{noir} shows the groups the four tags containing the word `noir,' `noir,' `film noir,' `neo noir' and `noir thriller,' belong to. It is interesting to see that `noir thriller' does not carry the exact same meaning as `noir' `film noir' or `neo noir' when applied by MovieLens users.

\newpage
Next, we build the feature-weighted `film profile' in a similar framework to that used by \citet{symeonidis2007feature} for constructing the feature-weighted user profile in a recommendation system. The `feature' in our context refers to the group of analogous tags. For each film, we can calculate the frequency of each tag group for it. However, such frequency can only describe how likely a film is related to a tag group, but not how much this tag group can differentiate this film from others. For instance, Martin Scorsese's period drama \textit{Age of Innocence} has the tag group \{based on a book, adapted from a book\}, with frequency 2. This tag group appears at least once for $1533$ films. Another tag group \{1800s\} for the film with frequency 1, appears at least once for only $11$ films. Thus \{1800s\} should be a more important  discriminating tag group than \{based on a book, adapted from a book\} for \textit{Age of Innocence}. But the frequency of the former is larger.

We prefer emphasizing a relatively rare tag group applied to a particular film that may potentially better distinguish it from others. Inspired by the TFIDF (term frequency–inverse document frequency) scheme in information retrieval \citep{baeza1999modern}, we define the inverse film frequency (IFmF) of the tag group $\text{Tg}$ as $\text{IFmF(Tg)}=\log\frac{N}{\text{FmF(Tg)}}$, where FmF(Tg) is the number of films for which tag group Tg appears at least once. From the formula, one sees that a tag group appearing for fewer films has a larger IFmF value. We define the multiplication of a tag group's frequency for a film and its IFmF as tag group frequency-inverse film frequency (TgFIFF), a value that gives more weight to uncommon tag groups. A feature-weighted vector of length $M=1043$ is then assigned to each film, in which each coordinate is the TgFIFF for it with respect to a tag group. By calculation, for \textit{Age of Innocence}, the TgFIFF of \{based on a book, adapted from a book\} is $6.222$ and that of \{1800s\} is $8.048$. Now \{1800s\} is weighted more. See Appendix \ref{fvec} for calculation details of the feature-weighted vectors.
\subsection{One-class nearest neighbors algorithm to identify noirish films}\label{oneclass}
The one-class nearest neighbors classification to identify noirish films will be performed using the feature-weighted vectors for films. It is worth noting that compared with one-class classification algorithms dealing with situations where one lacks negative samples, such samples are not rare in our test set, as there are plenty of films without noir attributes. 

The nearest neighbors in our case refer to the IMDb noirs in our training set $\mathcal{T}$ that are most similar to a film of interest in the unlabeled dataset $\mathcal{U}$. We normalize all the feature-weighted vectors for films to unit length, and the angle between any two of them divided by $\pi$ is a distance metric on a $(M-1)$-dimensional unit sphere. This distance measures two films' dissimilarity, and the nearest neighbors of an unlabeled films are the IMDb noirs having the shortest distance with it. A detailed illustration of our distance metric can be found in Appendix \ref{distance}.

However, there usually exist noisy observations in $\mathcal{T}$ in the real-world data collecting process \citep{khan2014x}. In particular, considering all IMDb noirs may be problematic due to the sparsity of tags. Some films in $\mathcal{T}$ only have one tag `photography,' which is generic, and any other film that also has this one single tag will have a zero distance from it. Some have very discriminating tag groups compared with others: Samuel Muller's noir \textit{House of Bamboo} is the nearest neighbor in $\mathcal{T}$ of Daniel Mann's comedy \textit{The Teahouse of the August Moon} in $\mathcal{U}$. They share three tag groups \{geisha, Japanese culture\}, \{Japan, Japanese, Tokyo, Yakuza\} and \{air force, army, military\}. By calculation, any other IMDb noirs are less close to \textit{House of Bamboo} in terms of angular distance than \textit{The Teahouse of the August Moon}. However, among $453$ films in $\mathcal{T}$, only $3$ have the tag group \{air force, army, military\}, $3$ have the tag group \{Japan, Japanese, Tokyo, Yakuza\} and \textit{House of Bamboo} is the only one having the tag group \{geisha, Japanese culture\}. So these tag groups might not be representative of $\mathcal{T}$ and \textit{House of Bamboo} might just be noise.

To identify noise, we modify the center-based distance method used by \citet{khan2018relationship}. For an unlabeled film $z$, we find its $J=3$ nearest neighbors $NN_j(z)$'s in the set of non-noise IMDb noirs, denoted by $\mathcal{S}$, and the nearest neighbor of $NN_j(z)$ in $\mathcal{S}$, $NN(NN_j(z))$, where $j=1,2,3$. Denoting the distance between two films $z_1$ and $z_2$ by $d(z_1,z_2)$, we calculate $d(z,NN_j(z))$ and $d(NN_j(z),NN(NN_j(z))$ and the ratio $r(z)$ of the mean of $d(z,NN_j(z)$'s to the mean of $d(NN_j(z),NN(NN_j(z))$'s. A threshold vector $\theta=(\theta_1,\theta_2,\theta_3,\theta_4)$ is used for the classification of $z$: The film $z$ is classified into $\mathcal{P}$, films with noir characteristics, if and only if $r(z)<\theta_1$ and $d(z,NN_j(z))<\theta_{j+1}$ for $j=1,2,3$ and this film has at least 5 tags. Films with fewer than 5 tags from the unlabeled set are excluded because we need a sufficient amount of information from the tagging. There are $99$ non-noise IMDb noirs in $\mathcal{S}$, and the process of noise recognition in Appendix is elaborated in Appendix \ref{noise}. The threshold vector chosen is $(1.26,0.43,0.43,0.43)$ through the cross-validation method described in Appendix \ref{thre}.

We develop the methodology on the basis of a popular scheme of one-class classification \citep{tax2002one}, which finds an unlabeled object $z$'s nearest neighbor in the training set, $NN(z)$, and the nearest neighbor(s) of $NN(z)$ in the training set, $NN(NN(z))$, and then considers the ratio of distance between $z$ and $NN(z)$ to the distance between $NN(z)$ and $NN(NN(z))$. If this ratio is smaller than a user-specified threshold, then we say $z$ is accepted, meaning it is classified into the positive class. Some alternative methods have been proposed by finding more than one nearest neighbor of $z$, or finding more than one nearest neighbor of $NN(z)$, or both, and then calculating either the ratio of the mean of distances or taking the majority vote to make the decision \citep{cabral2009combining, khan2018relationship}.

Our algorithm is different from the work of \citet{cabral2009combining} because we calculate the ratio of the means of distances, instead of calculating the ratios of distance between $z$ and $NN_j(z)$ and that between $NN_j(z)$ and $NN(NN_j(z))$ for each $j$ and then taking the majority vote. Moreover, our algorithm is different from the work of \citet{khan2018relationship} because we not only consider the ratio of mean distance, but also the distances themselves between the object and its nearest neighbors: We do not want to accept objects with a distance too far from its nearest neighbors in $\mathcal{S}$.

Our method to identify films with noir characteristics using users' tags, can be summarized as follows:\\
1) We clean the MovieLens and IMDb datasets and select $51,402$ narrative films of interest.\\
2) We pre-process the tagging data, extracting tag stems and focusing on relatively common tags with useful information for film classification. There are $34,412$ films and $2,788$ tags for analysis. \\
3) The pre-processed tags are clustered based on their occurrences for films to further address the fact that similar films may share similar but not exactly the same tags. Each cluster represents a tag group and each film is then assigned a feature-weighted vector.\\
4) Among the $34,412$ films, $453$ labeled as noir by IMDb form the training set and all the other films form the test set. The training set is partitioned into non-noise films and noises, and for a given film $z$ with at least 5 tags in the test set, we find the three nearest neighbors of it in the non-noise training subset,$NN_1(z)$, $NN_1(z)$,  $NN_3(z)$, and the nearest neighbor of $NN_j(z)$'s, $NN(NN_j(z))$, $j=1,2,3$. The film $z$ is classified into the positive class if and only if the ratio of the mean distance between $z$ and $NN_j(z)$ and the mean distance between $NN_j(z)$ and $NN(NN_j(z))$ is smaller than 1.26 and each $NN_j(z)$ is smaller than 0.43. These threshold values are chosen via applying the nearest neighbor approach to the training set with noises treated as negative proxies.
\section{Results and Discussion}\label{discussion}
There are $1,148$ films in the test set $\mathcal{U}$ with at least 5 tags that are classified into $\mathcal{P}$, the class of films with noir characteristics. The complete list of these films can be found in the \nameref*{sppl}. Table \ref{tableres} and Table \ref{tableres2} show some of them and their non-noise nearest neighbor in the set of non-noise IMDb noirs denoted by $\mathcal{S}$. The `Noir Tag' column indicates whether a film has at least one tag containing the word `noir,' with the value 1 meaning it does. Our algorithm is able to detect films' noirness even if they are not explicitly tagged with it. \footnote{Note that in the works cited by Table \ref{tableres}, \citet{selby1984dark}'s chronology of the film noir only includes films released between 1940 and 1959, and \citet{ward1979film}'s only includes films between 1927 and 1976.}
\begin{table}[h!]
\centering
\caption {Some of the films in $\mathcal{U}$, the unlabeled set classified into $\mathcal{P}$, films with noir characteristics. The `Noir Tag'' column is the value of the indicator function regarding whether a film has at least one of the tags `noir,' `film noir,' `neo noir' and `noir thriller,' and the abbreviations in the `Comment' column stand for various citations: BC for \citet{borde2002panorama}, N for \citet{naremore2008more}, S for \citet{selby1984dark}, and WS for \citet{ward1979film}.} 
\label{tableres}
\begin{tabular}{|c|c|c|c|c|}
\hline
Film&Country/Region&Year&Noir Tag&Comment \\
\hline
\textit{M}&Germany&1931&0&Expressionism film (BC, N)\\
\hline
\textit{The Public Enemy}&US&1931&0&Pre-code gangster film (WS)\\
\hline
\textit{La Bête Humaine}&France&1938&0&Poetic Realism noir (BC, N)\\
\hline
\textit{Hotel Reserve}&UK&1944&0&Noirish spy thriller (N)\\
\hline
\textit{Leave Her To Heaven}&US&1945&1&Color noir (BC, N, S, WS)\\
\hline
\textit{The Paradine Case}&US&1947&0&Noir (S)\\
\hline
\textit{Bob le Flambeur}&France&1956&0&Heist noir\\
\hline
\textit{Breathless}&France&1960&1&New Wave noir (N)\\
\hline
\textit{High and Low}&{Japan}&1963&1&Police crime noir (N)\\
\hline
\textit{A Fistful of Dollars}&Italy&1964&0&Spaghetti Western noir (N)\\
\hline
\textit{The Long Goodbye}&US&1973&1&Neo noir  (N, WS)\\
\hline
\textit{Insomnia}&Norway&1997&0&Neo noir (N)\\
\hline
\textit{Infernal Affairs}&HK, China&2002&0&Neo noir\\
\hline
\end{tabular}
\end{table}
\begin{table}[h!]
\centering
\caption {Nearest neighbor in the set of non-noise IMDb noirs $\mathcal{S}$ of films listed in Table \ref{tableres}}
\label{tableres2}
\begin{tabular}{|c|c|}
\hline
Film $z$ in $\mathcal{U}$&Film $z$'s nearest neighbor in  $\mathcal{S}$\\
\hline
\textit{M}&\textit{Touch of Evil}\\
\hline
\textit{The Public Enemy}&\textit{Fury}\\
\hline
\textit{La Bête Humaine}&\textit{The 39 Steps}\\
\hline
\textit{Hotel Reserve}&\textit{Ministry of Fear}\\
\hline
\textit{Leave Her To Heaven}&\textit{The Postman Always Rings Twice}\\
\hline
\textit{The Paradine Case}&\textit{Witness for the Prosecution}\\
\hline
\textit{Bob le Flambeur}&\textit{Murder, My Sweet}\\
\hline
\textit{Breathless}&\textit{Ace in the Hole}\\
\hline
\textit{High and Low}&\textit{Stray Dog}\\
\hline
\textit{A Fistful of Dollars}&\textit{Dead End}\\
\hline
\textit{The Long Goodbye}&\textit{Murder, My Sweet}\\
\hline
\textit{Insomnia}&\textit{Touch of Evil}\\
\hline
\textit{Infernal Affairs}&\textit{The Narrow Margin}\\
\hline
\end{tabular}
\end{table}

Our result evidences film noirs' intertexual connections with German Expressionism, French Poetic Realism, early British thrillers and American pre-code crime films. The influence of Expressionism on noir has been challenged by \citet{elsaesser1996german}, but it seems that the audience still tends to relate this modernist movement to noirs, as Lang's \textit{M} and \textit{The Testament of Dr. Mabuse} are the only films outside the US identified with noir attributes before 1935. The audience-recognized noirish French Poetic Realism films include Jean Renoir's \textit{La Bête Humaine}. Its 3 nearest neighbors include Lang's remake of it, \textit{Human Desire}, and Marcel Carné's \textit{Le Quai des brumes}, which are conceived as precursors to noir by some critics nowadays. The no-future pessimism has not been neglected by the audience, which is, as \citet{vincendeau1992noir} concludes, the `direct link between Poetic Realism and film noir.' Another European influence on the noirs observed from our result comes from British thrillers, among them are Alfred Hitchcock's \textit{Young and Innocent} and \textit{Foreign Correspondent} and Lance Comfort, Multz Greenbaum and Victor Hanbury's \textit{Hotel Reserve}. Although these films were ignored by the French critics in the early discussion of noir, with hindsight they made nonnegligible contributions to the noir category \citep{naremore2008more}. A number of Hitchcock's Hollywood thrillers are identified with noir attributes as well, suggesting that the Master of Suspense's noir style can be traced back to his British origin. Meanwhile, the American noirs may find their native roots in the pre-code crime pictures such as Howard Hawks' gangster film \textit{Scarface} and W. S. Van Dyke's \textit{The Thin Man}, or at least that is what the MovieLens audience indicates.

Regarding the classic noir era of the 1940s and 1950s, film critics sit on the fence deciding whether some films are actually noirs. These include Stahl's color thriller \textit{Leave Her To Heaven}, Hitchcock's courthouse drama \textit{The Paradine Case}, and Otto Preminger's melodrama \textit{Daisy Kenyon}. as well as sometimes-called `noir crossovers' like Robert Wise's science fiction film \textit{The Day the Earth Stood Still} and Fred Zinnemann's western \textit{High Noon}. IMDb says these are not noirs, but our result speaks for the audience to say they are as noirish as those IMDb-labeled ones. 

IMDb states it takes \textit{Touch of Evil} as the end of noir and hence automatically leaves out all neo noirs. Our classification method, on the other hand, effectively categorizes films after 1958. We found films widely described as neo noirs such as Robert Altman's \textit{The Long Goodbye}, David Lynch's \textit{Blue Velvet} and Carl Franklin's \textit{Devil in a Blue Dress}, along with some noirs' cousins such as Lynch's feature debut \textit{Eraserhead} that provoke debates regarding whether they are noirs themselves. Notably, our algorithm discovers a number of non-US productions that may have noir elements, from both the classic noir era and post 1960s: from French heist films to New Wave masterpieces, from Italian Spaghetti Western to gialli, and from Europe to Asia. Some Japanese films bear noir hues, and since the 1980s, neo noirs have played a significant role in Hong Kong cinema. The new century has seen the rise of South Korean neo noirs.

As one can see in Table \ref{tableres2}, the nearest neighbors approach helps us find the most similar non-noise IMDb-labeled noir for a given film. \textit{Hotel Reserve}'s nearest noir neighbor is Lang's \textit{Ministry of Fear}, indicating how British thrillers paved the way for American espionage noirs. Not surprisingly, Akira Kurosawa's \textit{High and Low}'s nearest neighbor is his earlier work \textit{Stray Dog}.

We now consider the 99 non-noise IMDb-labeled noirs and the 1148 films recognized by us as the films with noir characteristics and compare these works prior to and after 1960. Table \ref{tableres_com} gives the most frequent 5 tag groups for these films of two different periods respectively, and Table \ref{tableres_com2} the most frequent 5 tag groups for films with noir characteristics after 1960 that have never been applied to those prior to 1960. Here frequency of a tag group for a set of films is defined as the number of times this tag group occurs. While stories of revenge and murder as well as dark atmospheric tones are associated with noirs regardless of the era, neo noirs question justice with more displays of police corruption and also appear to be more erotic and violent. This is not surprising, as the decline of Motion Picture Production Code and the emergence of TV and home video have fundamentally changed the film industry. The Code used to refuse fairly explicit scenes as well as endings with no punishment for the villains, but after it lost its power in the 1960s, the moral ambiguity in noirs became evident: Don Siegel's \textit{Dirty Harry}  lambasts the police system, with the eponymous hero portrayed by Clint Eastwood throwing away his badge at the end; Coen brothers' \textit{Miller's Crossing} went further, in which the government and the police are controled by the gangsters. Moreover, post 1960s films had to compete with TV and face a younger and wider audience. The expanding home video market provided another path for films to earn profits. The technology evolution of TV and VHS eventually gave rise to erotic thrillers. While such productions' theater runs may have been limited by their explicit content, their performances in the video market can be guaranteed as long as they satisfy the audience's curiosity. As \citet{williams1993erotic} put it, `Their primary brief generally ensures that there is something interesting to watch, since they explore danger and sex in a format which is both thriller and skin flick.' 
\begin{table}[h!]
\centering
\caption {Most frequent 5 tag groups for films with noir characteristics prior to and after 1960}
\label{tableres_com}
\begin{tabular}{|c|c|c|}
\hline
Frequency Rank&Prior to 1960&After 1960\\
\hline
1&\specialcell{\{black market, noir, film noir, \\neo noir, privat detective\}}&\{murder, revenge\}\\
\hline
2&\specialcell{\{AFI 100 Thrills, black and white, \\classic, frightening, old, remade\}}&\specialcell{\{thriller, intrigue,\\ mystery, suspense\}}\\
\hline
3&\{murder, revenge\}&\specialcell{\{gangster, organized crime,\\ loser, mob, mafia, mobster\}}\\
\hline
4&\specialcell{\{thriller, intrigue,\\ mystery, suspense\}}&\specialcell{\{atmospheric, cinematography,\\dark, performance, stylized\}}\\
\hline
5&\specialcell{\{atmospheric, cinematography,\\dark, performance, stylized\}}&\specialcell{\{bad cop, police corruption,\\ curruption, dirty cop, justice, \\police brutality, political corruption\}}\\
\hline
\end{tabular}
\end{table}
\begin{table}[h!]
\centering
\caption {Most frequent 5 tag groups for films with noir characteristics after 1960 that have never applied to those prior to 1960}
\label{tableres_com2}
\begin{tabular}{|c|c|}
\hline
Frequency Rank&Tag group\\
\hline
1&\{nudity topless nudity topless notable\}\\
\hline
2&\{blood, gore, blood splatter, extreme violence, gory \}\\
\hline
3&\{nudity full frontal notable, NC 17, notable nudity, nudity full frontal\}\\
\hline
4&\{1970s, 70s, disco\}\\
\hline
5&\{Christian, religion, faith, miracle, religion\}\\
\hline
\end{tabular}
\end{table}

Some films classified into noirish films might seem surprising at the first glance, but are revealed to be deeply connected with noirs after more careful examination. Among them are Wilder's \textit{The Apartment} and impressively, David Lean's \textit{Brief Encounter}, the film that inspires it. Claiming \textit{The Apartment} is a shift away from Wilder's previous noir releases including \textit{Double Indemnity} may be specious. Coated with a romantic comedy flavor, it is pervaded by a sense of depression in the company offices full of desks and Jack Lemmon's claustrophobic apartment. Through set design, it pictures the industrial society which turns `workers into zombies or robots,' indebted to Lang's Expressionim classic \textit{Metropolis} as much as the noirs in 1940s and 1950s \citep{naremore2008more}. Furthermore, its plot featuring infidelity also reminds one of this recurring subject of classic noirs. \textit{The Apartment}'s three nearest neighbors are Nunnally Johnson's \textit{Black Widow}, Lang's \textit{Human Desire}, and Preminger's \textit{Laura}. It shares the tag groups \{adultery, infidelity, affair, extramarit affair, wife husband relationship\}, \{depress, suicide, suicide attempt\} and \{apartment, neighbor\} with at least one of them, verifying that these elements make it reminiscent of film noirs. \citet{naremore2008more} takes \textit{Brief Encounter} as an example of non-noir films that have `stylistic qualities usually described as noir.' Our result thus further addresses the question on what definitive noir traits are, and which of them are absent in a non-noir like \textit{Brief Encounter}.

The appearance of some obviously missclassified films is usually caused by some of their elements being similar to those of noirs. David O. Selznick's 1937 version of \textit{A Star Is Born}, for instance, is a story in the movie business like \textit{Sunset Boulevard}, takes place in Los Angeles like William Dieterle's \textit{The Turning Point}, and concerns the husband and wife relationship like Tay Garnett's \textit{The Postman Always Rings Twice}. These noirs are exactly its three nearest neighbors. 

Audiences do not particularly apply tags like `hard-boiled novels' to noirs, but simply say they are `adapted from a book.' This general tag connects noir with a couple of book adaptations, especially revenge stories like \textit{Hamlet} and \textit{Monte Cristo}. Films such as \textit{The Leopard Man} by Tourneur, \textit{Ossessione} by Luchino Visconti and \textit{Le Corbeau} by Henri-Georges Clouzot, often regarded as noirs by film scholars, are not classified into noirish films using our method mainly because they have few relevant tags. 

The tags also show that the audience cares more about the narrative rather than the visual aspects of films. Characters such as detective and femme fatale, elements of story such as murder, gangster, and affair, and locales such as New York and Los Angeles are the common tags applied to noirs, whereas visual motifs that often pertain to noir including imbalanced composition and the Dutch angle are hardly noticed by the audience. Some film terminologies do appear, though, examples of which include German Expressionism, but these are mostly recapitulative words one might find in a film's brief introduction online. Our method has thus limitations in that it classifies films mainly based upon their stories.

Further research directions would utilize more data sources and combine audience's tags with their online reviews to identify films with noir characteristics, in order to better comprehend the Internet audience's take on noirness.
\section*{Acknowledgments}
I would like to thank Prof Adam J. Rothman and Prof Galin L. Jones for their mentorship and Prof Robert B. Silberman for providing insightful discussions.
\section*{Supplementary Materials}\label{sppl}
The supplementary materials include the list of all films that are considered noirish with at least 5 tags using our algorithm and the R codes.
\theendnotes
\bibliographystyle{agsm}
\bibliography{ref}

\appendix
\nocitesec{*}
\begin{appendices}
\setcounter{equation}{0}
\setcounter{figure}{0}    
\renewcommand{\theequation}{\thesection.\arabic{equation}}
\section{Deriving films' feature-weighted vectors}
\subsection{Tag clustering algorithm}\label{clustering}
Denote the film-tag binary matrix as $\Gamma$, where the element on the $n$th row and $l$th column $\gamma_{nl}$ is 1 if tag $t_l$ is applied to film $f_n$, and 0 otherwise. The cosine similarity between any two tags $t_{l_1}$ and $t_{l_2}$ is defined as
\begin{equation}\nonumber
\cos(\gamma_{\cdot l_1},\gamma_{\cdot l_2})=\frac{<\gamma_{\cdot l_1},\gamma_{\cdot l_2}>}{|\gamma_{\cdot l_1}||\gamma_{\cdot l_2}|},
\end{equation}
where $\gamma_{\cdot l}$ is the $l$th column in $\Gamma$, and $<\cdot,\cdot>$ is the dot product of two vectors, $l_1,l_2=1,2,..., L$. The tag or tags with the highest cosine similarity to the $l$th tag will be considered as the strongly related tags of that. If tag $t_{l_1}$ is strongly related to tag $t_{l_2}$, we draw an edge from $t_{l_1}$ to $t_{l_2}$ with weight $w_{l_1,l_2}=\cos(\gamma_{\cdot l_1},\gamma_{\cdot l_2})$. In this way we build a directed graph, G, with $478$ connected subgraphs.

Clustering the directed graph would be difficult due to the asymmetry of the weight matrix \citep{malliaros2013clustering}. So, for each directed subgraph, we get the undirected subgraph by setting the edge weight between $t_{l_1}$ and $t_{l_2}$ as $w^{ud}=w_{l_1,l_2}+w_{l_2,l_1}$. Then we obtain the optimal partition by the maximizing-modularity algorithm proposed by \citet{brandes2007modularity}. This can be done in R using the function optimal\_clustering in the igraph package \citep{csardi2006}. We obtain $M=1043$ partitions in total for the $478$ connected subgraphs.
\subsection{Calculating the feature-weighted vectors for films}\label{fvec}
Denote the $L \times M$ tag-tag group binary matrix by $\Psi$, where the element on the $l$th row and $m$th column $\psi_{lm}$ is 1 if tag $t_l$ is in the group $\text{Tg}_{m}$, and $0$ otherwise. The film-tag group frequency matrix is then $\Lambda=\Gamma \Psi$. The element $\lambda_{nm}$ is the frequency of tag group $\text{Fe}_m$ for film $\text{Fm}_n$, or equivalently, the number of tags assigned to feature $\text{Tg}_m$ applied for this film.
We define a new matrix $\Phi$, in which the element $\phi_{nm}$ is the tag group frequency-inverse film frequency for film $\text{Fm}_n$ with respect to tag group $\text{Tg}_m$:
\begin{equation}\nonumber
    \phi_{nm}=\lambda_{nm}*\text{IFmF(Fe)})_m=\lambda_{nm}*\log\frac{N}{\text{FmF(Fe)}}.
\end{equation}
Then the $n$th row of $\Phi$ is the feature-weighted vector for film $\text{Fm}_n$.
\section{Deriving the one-class nearest neighbors algorithm}\label{mainalg}
\subsection{Defining the angular distance metric}\label{distance}
To define the distance metric for our algorithm, we first take the cosine similarity between any two rows of $\Psi$ as the similarity measure between the corresponding films. We can always normalize a row of $\Phi$ to make it a unit vector, so without the loss of similarity, we assume that all rows of $\Phi$ are unit vectors. Then the angle between two rows $\phi_{n_1,\cdot}$ and $\phi_{n_2,\cdot}$, $\text{arccos}(\cos(\phi_{n_1,\cdot},\phi_{n_2,\cdot}))/\pi$,
is a distance metric on a $(M-1)$-dimensional unit sphere. The angular distance between $\phi_{n_1,\cdot}$ and $\phi_{n_2,\cdot}$ is then defined by
\begin{equation}
\label{dis}
    \text{d}(\phi_{n_1,\cdot},\phi_{n_2,\cdot})=\text{arccos}(\cos(\phi_{n_1,\cdot},\phi_{n_2,\cdot}))/\pi.
\end{equation}
Since all coordinates of $\phi_{n,\cdot}$ are nonnegative, $\cos(\phi_{n_1,\cdot},\phi_{n_2,\cdot})\geq 0$ and thus $\text{d}(\phi_{n_1,\cdot},\phi_{n_2,\cdot}) \leq 0.5$. 
\subsection{Identifying noises in the training set of IMDb noirs}\label{noise}
Inspired by \citet{khan2018relationship}, we use a modified center-based distance method to identify noises in the training set $\mathcal{T}$. The center is the mean of normalized feature-weighted vectors in $\mathcal{T}$, and we normalize this center to make it a unit vector on the sphere. After calculating the distance defined by \eqref{dis} between each data point in $\mathcal{T}$ and the center, we get the quartiles of these distances, and consider any objects with a distance larger than the third quartile as noises in $\mathcal{T}$. Furthermore, films with fewer than 5 tags are also considered as noises since they can easily have high cosine similarities. In such case they will have very small distances to films in the unlabeled set that share few tags with them. Now we have 99 remaining positive non-noise objects (films) and 354 noises in the IMDb noir class. 
\subsection{Selecting the threshold vector}\label{thre}
Formally, the ratio $r(z)$ is defined as 
\begin{equation}
\label{rz}
r(z)=\frac{\sum_{j=1}^J \text{d} (z,NN_j(z))/J}{{\sum_{j=1}^J \text{d} (NN_j(z),NN(NN_j(z)))/J}}.
\end{equation}
We would want to find the threshold vector, $\theta=(\theta_1,\theta_2,...,\theta_J)$, so that a film $z$ with at least 5 tags is accepted if and only if $r(z)<\theta_1$ and $\text{d} (z,NN_j(z))<\theta_{j+1}$ for $j=1,2,...J$. Firstly, we reject any $z$ in $\mathcal{U}$ satisfying $\text{d} (z,NN_j(z))=0.5$, meaning that they do not share any tag groups with any objects in $\mathcal{S}$. 

Our process to select $\theta$ is motivated by the method to find the threshold of $r(z)$ when $J=1$ via cross-validation proposed by \citet{khan2018relationship}. The data in $\mathcal{T}$ are randomly split into $G$ folds such that each fold contains objects in both $\mathcal{S}$ and $\mathcal{T} \setminus \mathcal{S}$. The non-noises are treated as positive objects while the noises are treated as proxy for negative ones. For $g=1,2,...,G$, the $g$th fold is the validation set the one-class nearest neighbors algorithm is performed on using the non-noise objects in all the other $G-1$ folds as the training data. Choosing $J=3$, we select the first threshold value from the set of candidates
\begin{equation}
    \nonumber
\{\theta=(\theta_1,\theta_2,\theta_3,\theta_4)|\theta_1\in \{0.80,0.85,...,1.50\},\theta_2,\theta_3,\theta_4\in\{0.25,0.3,...,0.55\} \ \text{and}\ \theta_2\leq\theta_3 \leq \theta_4\}. 
\end{equation}
Since the distance between any two data points cannot exceed $0.5$, when $\theta_{j+1}=0.55$ it means there is no threshold for $\text{d} (z,NN_j(z))$ except for $\text{d} (z,NN_1(z))<0.5$ mentioned previously. Setting $G=5$, the 5-fold cross-validation is performed on $\mathcal{T}$. This procedure is repeated $100$ times and each time the candidate for $\theta$ giving the largest $\sqrt{\text{TPR}*\text{TNR}}$ is recorded. Here TPR is the true positive rate and TNR is the true negative rate, defined in the following equations
\begin{equation}\label{tpr}
\nonumber
    \text{TPR}=\frac{\sum_{g=1}^G \{\# i: r(z_{g_i})<\theta_1, \text{d} (z,NN_j(z))<\theta_{j+1} \ \text{for}\ j=1,2,3, \ \text{and}\ z_{g_i} \in \mathcal{S} \}}{|\mathcal{S}|},
\end{equation}
\begin{equation}\label{tfr}
\nonumber
     \text{TFR}=\frac{\sum_{g=1}^G \{\# i: r(z_{g_i}) \geq \theta_1\ \text{or}\ \text{d} (z,NN_j(z))\geq \theta_{j+1} \ \text{for some} \ j=1,2,3, \ \text{and} \ z_{g_i} \in \mathcal{T}\setminus\mathcal{S} \}}{|\mathcal{T}\setminus\mathcal{S}|},
\end{equation}
where $z_{g_i}$ is an object in the $g$th fold. Taking the majority vote, $(1.25,0.45,0.45,0.45)$ is chosen. Then we select the second threshold value from
\begin{equation}
    \nonumber
\{\theta=(\theta_1,\theta_2,\theta_3,\theta_4)|\theta_1\in \{1.20,1.21,...,1.30\},\theta_2,\theta_3^C,\theta_4^C\in\{0.40,0.41,...,0.50\} \ \text{and}\ \theta_2\leq\theta_3 \leq \theta_4\}
\end{equation}
and repeat the same procedure to finally choose $(1.26,0.43,0.43,0.43)$. 

We set $\theta=(1.26,0.43,0.43,0.43)$ as our final threshold. Since $\theta_{j+1}=0.43<0.5$ for all $j=1,2,3$, thresholding $\text{d} (z,NN_j(z))$ is giving a larger $\sqrt{\text{TPR}*\text{TNR}}$ than not thresholding them.
\end{appendices}

\bibliographystylesec{agsm}
\bibliographysec{refa}

\end{document}